\documentclass{article}
\usepackage{ijcai21}

\usepackage{times}

\usepackage{soul}
\usepackage{url}
\usepackage[hidelinks]{hyperref}
\usepackage[utf8]{inputenc}
\usepackage[small]{caption}
\usepackage{graphicx}
\usepackage{amsmath}
\usepackage{booktabs}
\usepackage{subcaption}
\usepackage{multirow}
\usepackage{comment}

\title{Optimal Team Economic Decisions in Counter-Strike}

\author{
Peter Xenopoulos$^1$\and
Bruno Coelho$^1$\and
Claudio Silva$^1$

\affiliations
$^1$New York University\\

\emails
\{xenopoulos, bruno.coelho, csilva\}@nyu.edu
}

\begin{document}

\maketitle

\begin{abstract}
The outputs of win probability models are often used to evaluate player actions. However, in some sports, such as the popular esport Counter-Strike, there exist important team-level decisions. For example, at the beginning of each round in a Counter-Strike game, teams decide how much of their in-game dollars to spend on equipment. Because the dollars are a scarce resource, different strategies have emerged concerning how teams should spend in particular situations. To assess team purchasing decisions in-game, we introduce a game-level win probability model to predict a team's chance of winning a game at the beginning of a given round. We consider features such as team scores, equipment, money, and spending decisions. Using our win probability model, we investigate optimal team spending decisions for important game scenarios. We identify a pattern of sub-optimal decision-making for CSGO teams. Finally, we introduce a metric, \textit{Optimal Spending Error} (OSE), to rank teams by how closely their spending decisions follow our predicted optimal spending decisions. 



\end{abstract}

\newcommand\blfootnote[1]{%
  \begingroup
  \renewcommand\thefootnote{}\footnote{#1}%
  \addtocounter{footnote}{-1}%
  \endgroup
}

\blfootnote{Accepted to IJCAI 2021 AISA Workshop}

\section{Introduction}
While traditional sports, such as basketball or soccer, have embraced data to analyze and value players, esports, also known as professional video gaming, still lags behind. Expanded data acquisition methods, such as player-worn sensors or cameras that track player movement, typically have driven analytical efforts in a sport. However, esports is unique in that many player actions and trajectories are recorded by a game server. In many esports, like Counter-Strike: Global Offensive (CSGO), a popular round-based first person shooter (FPS), game server logs are written to a \textit{demofile}. Because CSGO demofiles are easier to obtain and parse than demofiles from other esports, CSGO represents a promising direction for esports analytics, particularly in player and team valuation.

A common approach to valuing players is through the use of win probability models. As players perform actions in a CSGO round, they change their team's probability of winning that round. We can value players through the cumulative value of their actions~\cite{XenopoulosWPACSGO}. While evaluating players in CSGO is important for teams, fans and leagues alike, there exist many team-level actions that are important to value. For example, at the beginning of each CSGO round, two teams asymmetrically determine how much of their in-game money to spend on in-game equipment. Given the current game situation, such as the scores between the teams, as well as each team's current equipment value and money, a team will decide how much of their money to spend. 

Teams consider several different spending strategies depending on the game scenario. For example, some teams may elect to ``save'' for a round, meaning they limit how much in-game money they spend. On the other hand, some teams may ``half buy'', meaning they spend some of their money, but still retain some dollars for future rounds. The relative outcomes of the different strategies on win probability are unknown, since there are many factors to consider, such as team scores, equipment values and team money. Furthermore, current win probability models are tailored towards round-level win probability, whereas teams are making decisions with game-level win probability in mind.

In this paper, we introduce a game-level win probability model using features such as team scores, team equipment values and total team money. We consider various models, such as tree-based models and neural networks, and compare their performance to a logistic regression baseline. Then, we use these win probability models to assess the optimal spending decisions for various CSGO scenarios. We identify important game situations where the CSGO community is making sub-optimal decisions. Additionally, we introduce \textit{Optimal Spending Error (OSE)}, a metric to measure how much a team's spending decisions deviate from optimal ones.

We structure the paper as follows. In section~\ref{sec:rw}, we cover literature on win probability and valuation frameworks for esports. Then, in section~\ref{sec:csgo}, we describe CSGO as game, as well as how CSGO data is collected. Section~\ref{sec:win_prob} describes our data and win probability problem formulation, along with the performance of our candidate models. Next, in section~\ref{sec:discussion}, we discover the optimal spending decisions in various CSGO scenarios, as well as introduce our OSE metric. Finally, we discuss limitations, future work and conclude the paper in section~\ref{sec:conclusion}.
\section{Related Work} \label{sec:rw}
Predicting the chance that a team will win a game is an important task in sports analytics. The models used to predict who will win a game are typically called \textit{win probability} models, and often use the current state of a game as input. From these win probability models, we can assess the value of player actions and decisions. For example, Yurko~et~al.~value American football players by how their actions change their respective team's chance of winning the game~\cite{yurko2019nflwar}. Likewise, Decroos~et~al.~value soccer players by observing how their actions chance their team's chance of scoring~\cite{decroos2019}. Valuing actions and decisions through changes in a team's chance of winning and scoring is common among contemporary sports, such as in ice hockey~\cite{luo20,liu2018deep}, basketball~\cite{Sicilia19,cervone2014}, or soccer~\cite{fernandez2019decomposing}.

With new data acquisition pipelines, esports has also started to develop a win probability estimation and player valuation literature. Yang~et~al. built a model to predict the winning team in Defense of the Ancients 2 (DOTA2), a popular massively online battle arena (MOBA) esport video game~\cite{yang2016real}. They consider a combination of of pre-match features, as well as real-time features, as input to a logistic regression model. Semenov~et~al. use pre-match hero draft information to predict winners in DOTA2~\cite{DBLP:conf/aist/SemenovRKYN16}. Hodge~et~al. parse DOTA2 game replays, not only those from professional matches, but those of very high ranked public players, to predict outcomes in DOTA2~\cite{hodge2017}. Later, they also create a live win prediction model for DOTA2 using an ensemble of various machine learning techniques, such as logistic regression, random forests and gradient boosted trees~\cite{hodge2019win}. Recently, Kim~et~al. propose a confidence calibration method for predicting winners in League of Legends, a similar style game to DOTA2~\cite{Kim20}. Yang~et~al. introduce an interpretable model architecture to analyze win predictions in Honor of Kings, another popular MOBA game~\cite{yang2020}. Wang~et~al. provide a framework to jointly model win probability and player performance in Fever Basketball, a popular sports game~\cite{Wang20}.

While MOBA games have attracted increasing interest for win prediction and player valuation, first-person shooters (FPS), such as Counter-Strike, have received less attention. Makarov~et~al. predict round winners in CSGO in post-plant scenarios using decision trees and logistic regression~\cite{makarov2017predicting}. Bednarek~et~al. values players by clustering death locations, however, they do not create a win prediction model~\cite{bednarek2017player}. Recently, Xenopoulos~et~al. introduce a player valuation framework that uses changes in a team's win probability to value players~\cite{XenopoulosWPACSGO}. They model a team's chance of winning a CSGO round using XGBoost with input features such as a team's remaining players and total equipment. Prior CSGO work has focused on round-level win probability and player-level analysis. Our work differs in that we predict game-level win probability, and we analyze team decisions and actions, rather than those from players.
\section{Counter-Strike} \label{sec:csgo}
\subsection{Game Rules}

\begin{figure}
    \centering
    \includegraphics[width=\linewidth]{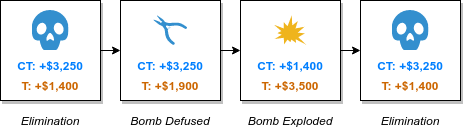}
    \caption{Teams earn \$3,250 per player for winning a round, or \$3,500 if they successfully detonate the bomb. When a team loses a round, they gain a variable amount of money (``loss bonus'') that depends on how many rounds they have lost since they last won a round. After each win, the loss bonus resets. Because the money from losing can be small, teams often find themselves in situations where they must determine how much money to invest to maximize their chance of winning the game.}
    \label{fig:loss_bonus}
\end{figure}

Counter-Strike: Global Offensive (CSGO) is the latest rendition in the popular Counter-Strike series of video games. In a professional CSGO match, two teams first meet to decide what \textit{maps}, or virtual worlds, they plan on playing. Usually, games are structured as a best-of-three, and the two teams go through a picking and banning process to choose maps they want to play. There are seven maps used in professional play, and the pool of seven maps changes occasionally. Teams win a map when they win 16 rounds, unless the score is 15 for each team; then, an overtime process takes place.

When teams begin to play a map, they are first assigned to either the Terrorists (T) or the Counter-Terrorists (CT). They then play for 15 rounds as their assigned side, and the two teams switch sides after 15 rounds. Each round lasts slightly under two minutes. Teams can achieve a variety of round win conditions based on their current side. The T side can win a round by eliminating all CT players, or by planting a bomb and having it defuse at one of two designated bomb sites. At the beginning of each round, a random T player starts with the bomb. When the bomb is planted, a timer for 35 seconds starts counting down. The CT side can win a round by eliminating all T players, defusing a planted bomb, or allowing the round time to run out, without reaching any of the aforementioned win conditions.

Players begin a round with 100 health points (HP) and are eliminated when their HP reaches zero. Players lose HP when they are damaged by gunfire or grenades from other players. At the beginning of each round, players can use money earned from previous rounds to buy guns, armor, grenades and other equipment. Players earn money by eliminating other players and through completing objectives. If a team wins a round, each member gains \$3,250 in-game money. However, if a team loses, their monetary gain is based on how many previous rounds they have lost since their last win. For example, if a team lost the previous round, but won two rounds ago, they only earn \$1,400. We show an example of the variable ``loss bonus'' in Figure~\ref{fig:loss_bonus}. Since the loss bonus can be minimal, teams often use strategies, known as \textit{buy types}, to guide their in-game economic decisions. Some buy types include deciding not to buy any equipment in a round (referred to as an ``eco''), or buying cheap guns (``low buy'' or ``half buy'') in an effort to save money while maximizing their chance of winning the game. If a player saves equipment from a previous round, but the team elects to either eco or half buy, they these are referred to as ``hero'' buys. Finally, if a team's starting equipment value, plus their round monetary spend, is \$20,000 or greater, the team's buy is referred to as a full buy. We define the main buy types in Table~\ref{tab:buy_strats}.

\begin{table}[]
\centering
\begin{tabular}{@{}rccc@{}}
\toprule
\multicolumn{1}{c}{\textbf{Buy Type}} & \textbf{Equipment} & \textbf{Spend} & \textbf{Win \%} \\ \midrule
Eco & 0 -- 3k & 0 -- 2k & 3\% \\
Low Buy & 0 -- 3k & 2k -- 7.5k & 27\% \\
Half Buy & 0 -- 3k & 7.5k -- 20k & 34\% \\
Hero Low Buy & 3k -- 20k & 0 -- 7.5k & 25\% \\
Hero Half Buy & 3k -- 20k & 7.5k -- 17k & 48\% \\
Full Buy & \multicolumn{2}{c}{Equip. $+$ Spend $>$ 20k+} & 59\% \\ \bottomrule
\end{tabular}
\caption{Teams use different buying strategies depending on their current money, and which strategy they think will improve their chance of winning the game. We also report the average round win rates for each buy type.}
\label{tab:buy_strats}
\end{table}

\subsection{Data Acquisition}
CSGO games take place on a game server, to which the clients (players) connect. Each client persists a local game state, containing information such as where other players are standing, team scores and so on. When a client performs some input that changes the game, that input is sent to the game server which then reconciles the global game state across all connected clients. As the game server updates the global game state, and sends it to each client, the game server also writes the change to a \textit{demofile}~\cite{bednarek2017data}. Demofiles are limited to a single map, so for a best-of-three game, at minimum, two demofiles are generated. One can parse a demofile into a JSON format using a demo parser~\cite{XenopoulosWPACSGO}.

\section{Modeling Win Probability} \label{sec:win_prob}

\begin{figure}
    \centering
    \includegraphics[width=\linewidth]{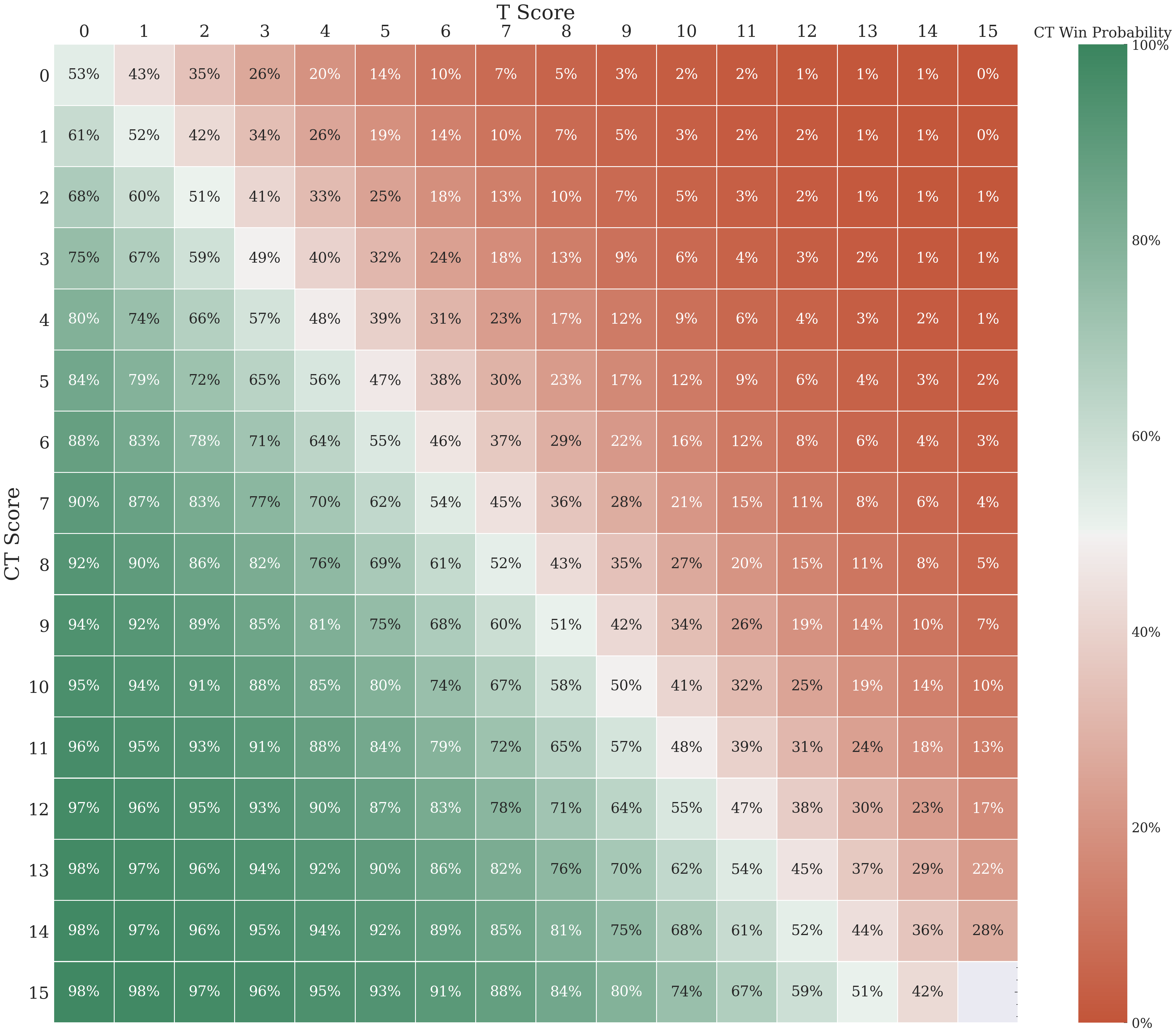}
    \caption{CT win probability by score on the map \textit{Inferno} from our baseline logistic regression model that uses team scores. Winning the first two rounds can increase a team's chance of winning the game by roughly 15\%.}
    \label{fig:inferno-win-rate}
\end{figure}

In previous win probability works, the objective is typically a regression problem starting with a game state $G_{r,t}$, which encapsulates the game information, such as player locations, equipment values and so on, in round $r$ at time $t$~\cite{XenopoulosWPACSGO}. Using this aforementioned game state, the goal is to estimate $P(Y_{r} = 1 | G_{r,t})$, where $Y_{r}$ is 1 if the CT side wins round $r$ and 0 otherwise. We vary from this formulation by instead considering $R_{g,i}$, which is the game information at the start of round $i$ for game $g$. We can represent $R_{g,i}$ as a vector including each side's current score, the difference in score, each side's starting equipment value and money, and the buy decisions for both sides in round $i$. Using $R_{g,i}$, we want to estimate $P(Y_{g} | R_{g,i})$, where $Y_{g}$ represents the game outcome decision for game $g$. $Y_g$ can represent three unique values: (1) the game is won by the current CT side, (2) the current T side, or (3) the game ends in a draw.

Past win probability works have considered a variety of models. For example, logistic regression and gradient boosted trees are often used to model win probability, particularly because they provide good performance and allow for easy interpretability measures~\cite{makarov2017predicting,decroos2019,yang2016real,DBLP:conf/aist/SemenovRKYN16}. At the same time, neural network approaches are relatively untested for win probability prediction, particularly in esports. Therefore, we consider three modeling techniques: (1) logistic regression, (2) XGBoost, and (3) neural networks~\cite{DBLP:conf/kdd/ChenG16}. Logistic regression serves as our baseline model, and we show the predicted game win probabilities given different score combinations on the \textit{Inferno} map in Figure~\ref{fig:inferno-win-rate}. We use the default parameters for logistic regression and fine tune the parameters of the other model as described in Section~\ref{subsec:hyperparam}. We train a separate model for each of the seven competitive maps in CSGO, as well as a model where we one-hot encode the map as a feature, denoted ``\textit{OHE Map}''. We evaluate each model by using log-loss, defined as

\begin{equation}
    -\frac{1}{N} \sum_{i=0}^{N-1} \sum_{k=0}^{K-1} y_{i,k} \log p_{i,k}
\end{equation}

\noindent where $K$ is the number of labels, $N$ is the number of observations. Each model's log-loss on the test set is shown in Table~\ref{tab:results}. While more complex models, like XGBoost and neural networks, clearly outperform our logistic regression baseline, the performance differences between XGBoost and our neural network are minimal. We found a small performance benefit to using one-hot encoding for the map feature.

\begin{table}[]
\centering
\begin{tabular}{@{}rcccc@{}}
\toprule
\multicolumn{1}{c}{\textbf{Map}} & \textbf{Rounds} & \textbf{\begin{tabular}[c]{@{}c@{}}Logistic\\ Regression\end{tabular}} & \textbf{XGBoost} & \textbf{\begin{tabular}[c]{@{}c@{}}Neural\\ Network\end{tabular}} \\ \midrule
dust2    &  8144   & 0.820 & \textbf{0.766}           & 0.767 \\

inferno  &  10508  & 0.769 & \textbf{0.708}  & 0.709 \\

mirage   &  7712   & 0.876 & \textbf{0.827}  & 0.830  \\

nuke     &  8266   & 0.773 & \textbf{0.716}  & \textbf{0.716} \\

overpass &  6230   & 0.717 &         0.673   & \textbf{0.660} \\

train    &  7565   & 0.824 &         0.767   & \textbf{0.766} \\

vertigo  &  6080   & 0.768 &         0.710  & \textbf{0.707} \\ \bottomrule

\multicolumn{2}{c}{\textit{Weighted Average}} & \textit{0.793} & \textit{0.739} & \textit{0.736} \\ 

\multicolumn{2}{c}{\textit{OHE Map}} & - & \textit{\textbf{0.730}} & \textit{0.732} \\ 
\end{tabular}
\caption{Log-loss results by map for each of our candidate models. The logistic regression baseline uses only team scores. The tuned XGBoost model and the tuned neural network have similar results, with the neural network having a slightly smaller weighted average loss.
Using map as a feature, instead of a separate model for each map, performs slightly better. 
}
\label{tab:results}
\end{table}

\subsection{Data} \label{subsec:data}
We use 6,538 demofiles from matches between April 1st, 2020 to April 20th, 2021. We use this time period as all matches were played online, however, for matches before April 2020, it was common for matches to be played on local area networks, where players were in the same room to lower game server ping. Liu~et~al.~find that lower latencies can drastically affect player performance~\cite{liu2021lower}. Therefore, matches held in LAN tournaments versus those held online may not be directly comparable. We acquired our demofiles from the popular CSGO fan site, HLTV.org. For training, we use matches from April 1st, 2020 to the end of September 2020 (3,308 demofiles), for validation, we use matches from October and November 2020 (1,108 demofiles), and finally, for testing, we use matches from December 1st, 2020 to April 20th, 2021 (2,122 demofiles).

\subsection{Hyperparameter Tuning} \label{subsec:hyperparam}
For XGBoost, we use the Hyperopt~\cite{hyperopt} optimization library with default parameters to fine-tune the learning rate, maximum tree depth, minimum child weight and the subsample ratio of columns for each level.
We use the validation set indicated in Section~\ref{subsec:data}.
On the final model we use early stopping on the validation loss with a patience of 10 iterations to also tune the number of boosting rounds.

For the neural network, we create a two-layer fully-connected network, where each layer has a ReLU activation function and dropout applied~\cite{DBLP:journals/jmlr/SrivastavaHKSS14}.
We base our two-layer design choice on the work of Yurko~et~al.~, where they considered a two layer fully-connected network to predict an American football ball carrier's end position as one of their candidate models~\cite{yurko2020going}.
We fine-tune the amount of hidden units in each layer, the dropout probability and the learning rate using random search and use early stopping with a patience of 10 epochs on the final model.
Our output layer contains three units, one for a CT win, T win and draw, the three possible game outcomes, to which we apply a softmax activation function.
In Table~\ref{tab:hyperparam-range} we specify the range explored for each hyperparameter. 
We use the same range for both tuning a model per map and for the OHE Map models.

\begin{table}[]
\centering
\begin{tabular}{@{}r|cccc@{}}
\toprule
\multicolumn{1}{c}{\textbf{Model}} & \textbf{Hyperparameter} & \textbf{Lower} & \textbf{Upper} & \textbf{Step} \\ \hline
\multirow{4}{*}{XGBoost} & Learning rate    & 0.05      & 0.31       & 0.05 \\
                         & Max depth        & 3         & 10         & 1    \\
                         & Min. child weight & 1         & 8          & 1    \\
                         & Colsample        & 0.3       & 0.8        & 0.1  \\
\hline                                    
\multirow{4}{*}{NN}      & Dense 1          & 16        & 128        & 16   \\
                         & Dense 2          & 16        & 128        & 16   \\
                         & Dropout          & 0.1       & 0.6        & 0.1  \\
                         & Learning rate               & $10^{-3}$ & $10^{-5}$  & 0.1*\\ 
\bottomrule
\end{tabular}
\caption{Range of values and step sizes considered for each hyperparameter. For the learning rate of the neural network, we instead use a exponential step increment and multiply each value by 0.1.}
\label{tab:hyperparam-range}
\end{table}

\section{Discussion} \label{sec:discussion}

\subsection{Investigating Common Buy Strategies}

\begin{table*}[]
\centering
\begin{tabular}{cccccccc}
\multicolumn{1}{l}{} & \multicolumn{1}{l}{} & \multicolumn{6}{c}{\textbf{ACTUAL}} \\ \cline{2-8} 
\multicolumn{1}{c|}{\multirow{6}{*}{\textbf{\begin{tabular}[c]{@{}c@{}}O\\ P\\ T\\ I\\ M\\ A\\ L\end{tabular}}}} & \multicolumn{1}{c|}{\textbf{CT}} & \textit{Eco} & \textit{Low Buy} & \textit{Half Buy} & \textit{Hero Low Buy} & \textit{Hero Half Buy} & \multicolumn{1}{c|}{\textit{Full Buy}} \\ \cline{2-8} 
\multicolumn{1}{c|}{} & \multicolumn{1}{c|}{\textit{Eco}} & 47 & 136 & 141 & 0 & 0 & \multicolumn{1}{c|}{756} \\
\multicolumn{1}{c|}{} & \multicolumn{1}{c|}{\textit{Low Buy}} & 258 & 270 & 1218 & 0 & 0 & \multicolumn{1}{c|}{2880} \\
\multicolumn{1}{c|}{} & \multicolumn{1}{c|}{\textit{Half Buy}} & 1675 & 2126 & 3557 & 0 & 0 & \multicolumn{1}{c|}{2173} \\
\multicolumn{1}{c|}{} & \multicolumn{1}{c|}{\textit{Hero Low Buy}} & 0 & 0 & 0 & 680 & 86 & \multicolumn{1}{c|}{1081} \\
\multicolumn{1}{c|}{} & \multicolumn{1}{c|}{\textit{Hero Half Buy}} & 0 & 0 & 0 & 1926 & 568 & \multicolumn{1}{c|}{4453} \\
\multicolumn{1}{c|}{} & \multicolumn{1}{c|}{\textit{Full Buy}} & 3 & 86 & 114 & 31 & 29 & \multicolumn{1}{c|}{25967} \\ \cline{2-8} 
\multicolumn{1}{l}{} & \multicolumn{1}{l}{} & \multicolumn{1}{l}{} & \multicolumn{1}{l}{} & \multicolumn{1}{l}{} & \multicolumn{1}{l}{} & \multicolumn{1}{l}{} & \multicolumn{1}{l}{} \\
\multicolumn{1}{l}{} & \multicolumn{1}{l}{} & \multicolumn{6}{c}{\textbf{ACTUAL}} \\ \cline{2-8} 
\multicolumn{1}{c|}{\multirow{6}{*}{\textbf{\begin{tabular}[c]{@{}c@{}}O\\ P\\ T\\ I\\ M\\ A\\ L\end{tabular}}}} & \multicolumn{1}{c|}{\textbf{T}} & \textit{Eco} & \textit{Low Buy} & \textit{Half Buy} & \textit{Hero Low Buy} & \textit{Hero Half Buy} & \multicolumn{1}{c|}{\textit{Full Buy}} \\ \cline{2-8} 
\multicolumn{1}{c|}{} & \multicolumn{1}{c|}{\textit{Eco}} & 20 & 23 & 26 & 0 & 0 & \multicolumn{1}{c|}{1} \\
\multicolumn{1}{c|}{} & \multicolumn{1}{c|}{\textit{Low Buy}} & 174 & 706 & 1647 & 0 & 0 & \multicolumn{1}{c|}{1378} \\
\multicolumn{1}{c|}{} & \multicolumn{1}{c|}{\textit{Half Buy}} & 1582 & 2283 & 4726 & 0 & 0 & \multicolumn{1}{c|}{157} \\
\multicolumn{1}{c|}{} & \multicolumn{1}{c|}{\textit{Hero Low Buy}} & 0 & 0 & 0 & 159 & 247 & \multicolumn{1}{c|}{1794} \\
\multicolumn{1}{c|}{} & \multicolumn{1}{c|}{\textit{Hero Half Buy}} & 0 & 0 & 0 & 241 & 196 & \multicolumn{1}{c|}{4225} \\
\multicolumn{1}{l|}{} & \multicolumn{1}{c|}{\textit{Full Buy}} & 0 & 63 & 623 & 30 & 13 & \multicolumn{1}{c|}{29946} \\ \cline{2-8} 
\end{tabular}
\caption{Confusion matrices of predicted optimal side buy types and actual side buy types for our test set. We see that in many instances, the predicted optimal buy type differs from the actual buy type, particularly for eco and low buy situations.}
\label{tab:confusion-matrix}
\end{table*}

\begin{figure}
    \centering
    \includegraphics[width=\linewidth]{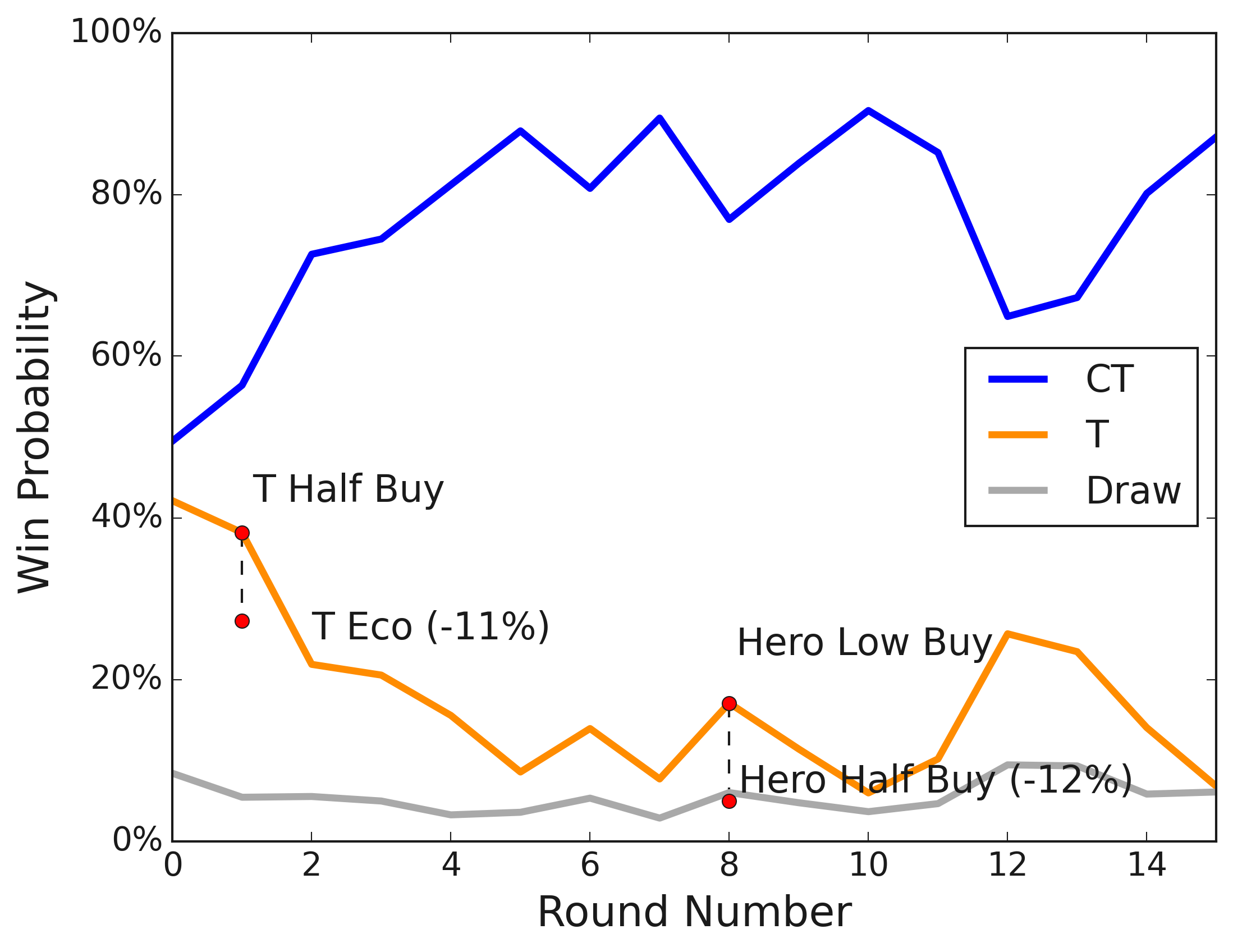}
    \caption{The estimated win probability for each side in the first half of a game on the \textit{Inferno} map. We show two alternate buys, one where buying \textit{less} (Eco instead of Half Buy) would have decreased the T's chance of winning, and one where buying \textit{more} (Hero Half Buy instead of Hero Low Buy) would have decreased the T's chance of winning.}
    \label{fig:example_game}
\end{figure}

One benefit of our approach is that we can estimate the game win probability for a variety of possible buy types. We would expect that different buys should naturally change a team's probability of winning a game. We show the effects of different buys on predicted team win probability in Figure~\ref{fig:example_game}. In doing so, we can determine the optimal buy type for a side in a given round by observing which buy type maximizes a team's game win probability. We present a confusion matrix in Table~\ref{tab:confusion-matrix} which shows the optimal buy types, along with the actual performed buy types for each side. While some buy types are overwhelmingly taken when optimal, such as the Full Buy, others, like an Eco buy, are performed more often than is predicted optimal. For example, out of the nearly 2,000 rounds where CT side performed an Eco buy, the predicted optimal buy was either a Low or Half buy, in over 90\% of rounds, leading to an average lost win probability of about 3\%. 

The first few rounds of any CSGO game are crucial for both sides. As we see from Figure~\ref{fig:inferno-win-rate}, by winning the first two rounds, the CT side achieves a 68\% game win rate on the \textit{Inferno} map. Such a high game win rate for winning the first two rounds is standard across all maps. The side which loses the first round must then decide what their buy type will be for the second round. Almost always, the possible buys will be an eco, low buy or half buy. The rational behind half buying is that the team which won the first round will still not have the best equipment available, thus by half buying, they will have a better chance of winning the second round and forcing their opponents to have a low loss bonus. However, if a team chooses to eco, they may save more money to spend in the third or fourth rounds. In Table~\ref{tab:second-round-buys}, we show the buy rates by side for second rounds where a team lost the first round. It is clear that in general, opting to half buy in the second round is a popular strategy, but in almost 40\% of rounds, the T side will either eco or low buy. At the same time, our model overwhelmingly predicts that a team should half buy in the second if they lose the first.

\begin{table}[]
\centering
\begin{tabular}{r|cccc}
\toprule
\multicolumn{1}{c}{} &  & \textbf{Eco} & \begin{tabular}[c]{@{}c@{}}\textbf{Low}\\ \textbf{Buy}\end{tabular} & 
\begin{tabular}[c]{@{}c@{}}\textbf{Half}\\ \textbf{Buy}\end{tabular} \\ \hline
\multirow{2}{*}{Actual} & T & 27\% & 10\% & 63\% \\
 & CT & 6\% & 3\% & 91\% \\ \hline
\multirow{2}{*}{Optimal} & T & 0\% & 0\% & 100\% \\
 & CT & 4\% & 0\% & 96\% \\ \hline
\end{tabular}
\caption{Actual and optimal buy type rates, by side, for second rounds where the team lost the first (pistol) round. Our model predicts that half buy is the most optimal buy type in most situations, increasing }
\label{tab:second-round-buys}
\end{table}

\subsection{Assessing Team Economic Decisions}

\begin{figure}
    \centering
    \includegraphics[scale=0.4]{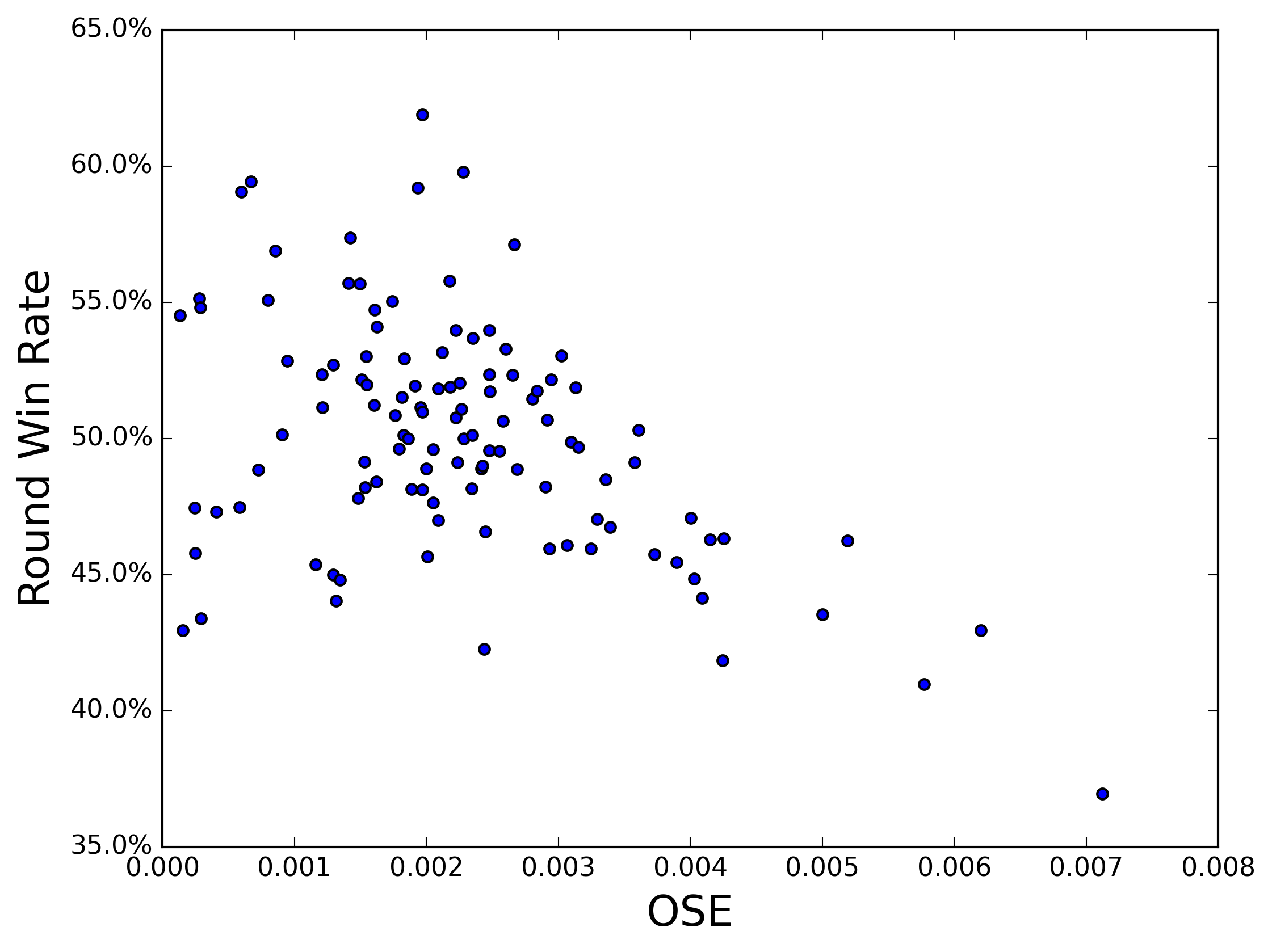}
    \caption{Teams with a high Optimal Spending Error (OSE), which translates into more sub-optimal team economic decision making, typically have lower round win rates ($r = -0.45$).}
    \label{fig:ose-rwr}
\end{figure}

Aside from observing community-wide team buy tendencies, we can also rank teams by how well their buys correlate with the optimal ones determined from our model. To that effect, we introduce \textit{Optimal Spending Error}, defined as

\begin{equation}
    OSE_T = \sum_{r \in R_T} (W_{T,r} - O_{T,r})^2
\end{equation}

\noindent which is the mean-squared error between the predicted win probability, and the optimal buy win probability, for team $T$ across all rounds which team $T$ plays, denoted $R_T$. Thus, teams with low OSEs are making economic decisions in line with what our model predicts as optimal. We calculate OSE for all teams in our test set, and report the results for teams ranked in the top 50 in Table~\ref{tab:ose-teams}. We see that in general teams with a top HLTV ranks have low OSEs, although the relationship is not completely captured by HLTV ranks, since teams experience different tiers of competition not delineated in the data. For example, a top 50 team is not often playing against a team ranked below the top 50. Therefore, we consider the relationship between a team's round win rate and their OSE. We would expect that teams which attain low OSEs, and therefore make optimal buy decisions, would also have high round win rates. We confirm this relationship between team OSE and round win rates in Figure~\ref{fig:ose-rwr}. We find a slight $(r = -0.45)$ relationship between team OSE and round win rate, indicating that teams with higher OSEs have lower round win rates. 

\begin{table*}[]
\centering
\begin{tabular}{ccllccccllc}
\multicolumn{5}{c}{\textit{Top 5}} &  & \multicolumn{5}{c}{\textit{Bottom 5}} \\ \cline{1-5} \cline{7-11} 
\multicolumn{1}{c}{\textbf{Team}} & \textbf{\begin{tabular}[c]{@{}c@{}}Average\\ OSE\end{tabular}} & \multicolumn{1}{c}{\textbf{\begin{tabular}[c]{@{}c@{}}CT\\ OSE\end{tabular}}} & \multicolumn{1}{c}{\textbf{\begin{tabular}[c]{@{}c@{}}T\\ OSE\end{tabular}}} & \multicolumn{1}{c}{\textbf{\begin{tabular}[c]{@{}c@{}}HLTV\\ Rank\end{tabular}}} & \multicolumn{1}{c}{} & \textbf{Team} & \textbf{\begin{tabular}[c]{@{}c@{}}Average\\ OSE\end{tabular}} & \multicolumn{1}{c}{\textbf{\begin{tabular}[c]{@{}c@{}}CT\\ OSE\end{tabular}}} & \multicolumn{1}{c}{\textbf{\begin{tabular}[c]{@{}c@{}}T\\ OSE\end{tabular}}} & \multicolumn{1}{c}{\textbf{\begin{tabular}[c]{@{}c@{}}HLTV\\ Rank\end{tabular}}} \\ \cline{1-5} \cline{7-11} 
\multicolumn{1}{c}{FPX} & 0.00024 & 0.00043 & 0.00005 & \multicolumn{1}{c}{16}                          & \multicolumn{1}{c}{} & ENCE & 0.00352 & 0.00501 & 0.00178 & \multicolumn{1}{c}{26} \\
\multicolumn{1}{c}{Gambit} & 0.00133 & 0.00090 & 0.00114 & \multicolumn{1}{c}{1}                       & \multicolumn{1}{c}{} & HAVU & 0.00309 & 0.00348 & 0.00271 & \multicolumn{1}{c}{14} \\
\multicolumn{1}{c}{SKADE} & 0.00141 & 0.00114 & 0.00165 & \multicolumn{1}{c}{23}                       & \multicolumn{1}{c}{} & Liquid & 0.00306 & 0.00312 & 0.00300 & \multicolumn{1}{c}{8} \\
\multicolumn{1}{c}{Dignitas} & 0.00160 & 0.00143 & 0.00176 & \multicolumn{1}{c}{25}                    & \multicolumn{1}{c}{} & \begin{tabular}[c]{@{}c@{}}Copenhagen\\ Flames\end{tabular} & 0.00265 & 0.00387 & 0.00155 & \multicolumn{1}{c}{33} \\
\multicolumn{1}{c}{NIP} & 0.00176 & 0.00122 & 0.00230 & \multicolumn{1}{c}{13}                         & \multicolumn{1}{c}{} & Anonymo & 0.00247 & 0.00264 & 0.00229 & \multicolumn{1}{c}{47} \\ \cline{1-5} \cline{7-11} 

\end{tabular}
\caption{Top and Bottom 5 teams by OSE. Only teams in the Top 50 HLTV.org rankings on April 26th, 2021, and have played at least 300 rounds in our test set, are included.}
\label{tab:ose-teams}
\end{table*}

\subsection{Accounting for New Maps}
\label{subsec:new-maps}


One interesting consideration is understanding how our models behave when a new map is added to the Active Duty Group, which is the list of active competitive maps.
To study the generalization abilities of our models, we train a neural network with no map features on six out of the seven available maps, and then evaluate the test loss on the held out map, imitating a scenario where a brand new map is introduced.

We maintain the neural network architecture described in section~\ref{subsec:hyperparam} and use 96 and 32 neurons for the first and second layer respective, 0.1 dropout probability and a learning rate of $10^{-4}$ with no hyperparameter tuning.
We then fine tune our model with a lower learning rate of $10^{-5}$ on the new map and observe the new test loss. Table~\ref{tab:transfer-learning} shows the loss for each map, where we also repeat the neural network results from Table~\ref{tab:results} for easier comparison.
We observe that both the initial models and the fine tuned models perform better or equal to our previous neural network trained on a specific map only, showing that for our problem the amount of data available for the model is more helpful than explicitly capturing effects of a specific map. 
This agrees with our previous results in Table~\ref{tab:results} where the OHE model had a lower average loss than the individual per map models across both neural networks and XGBoost. Thus, for future additions to the map pool, we can consider a model which uses no map information to achieve a rough win probability estimate.

\begin{table}[]
\centering
\begin{tabular}{@{}rccc@{}}
\toprule
\multicolumn{1}{c}{\textbf{Held out map}} & \textbf{\begin{tabular}[c]{@{}c@{}}Previous\\ best NN\end{tabular}} & \textbf{Initial model} & \textbf{Fine Tuned} \\ \midrule
dust2    & 0.767    &            \textbf{0.759} & 0.761               \\
              
inferno  & 0.709    &            0.715          & \textbf{0.705}     \\
             
mirage   & 0.830   &           \textbf{0.821} & 0.822               \\
              
nuke     & \textbf{0.716}   &            0.718        & \textbf{0.716}       \\
              
overpass & 0.660   &            \textbf{0.646} &         \textbf{0.646} \\
              
train    & 0.766   &            0.768 &         \textbf{0.759} \\
            
vertigo  & 0.707 &            0.697 &         \textbf{0.695} \\ \bottomrule
\end{tabular}
\caption{Log-loss results by map for our previous hyperparameter tuned models, a initial model which was trained with data from all other maps and one that was fine tuned on the held out map. The maps trained on more data, either before or after fine-tuning perform better in 6 out of 7 cases indicating there is little advantage to training a model on a specific map only.}
\label{tab:transfer-learning}
\end{table}
\section{Conclusion} \label{sec:conclusion}
This work introduces a game-level win probability model for CSGO. We consider features at the start of each round, such as the map, team scores, equipment, money, and buy types. We consider a variety of modeling design choices, from model architecture to how to encode map information. We then investigate team buying decisions across the professional CSGO community. We find that teams are much more conservative with respect to round buying decisions than our model predicts is optimal. Particularly in second rounds, where a team lost the first round of the game, many teams are making sub-optimal buy decisions on the T side, with respect to our model. We also introduce \textit{Optimal Spending Error} (OSE), a metric to assess a team's economic decision making. We find that OSE is correlated with other team success measures, such as round win rate. Finally, we conduct a test to assess how our model performs for ``unplayed'' maps, or maps that are not part of the active map pool. We find that our models can easily generalize to win probability prediction on new maps.

We see many avenues of future work. One of the limitations of our work is that our data is limited to organized semi-professional and professional games. Because these games often have teams with dedicated, oftentimes very similar, strategies for particular scenarios, in some cases there may not be much variation buy types. Consider the example in Table~\ref{tab:second-round-buys}. While there is variation in T buys for second rounds, the CT round buys have a high imbalance. One way to alleviate this issue is to also consider data from amateur, unorganized matches, which typically display more variation in buys. As buy decisions also exist in other FPS esports, such as Valorant, future work will be directed towards extending our framework to other games.

\bibliographystyle{named}
\bibliography{00_references}

\end{document}